\documentclass[sigconf]{acmart}
\acmConference[Anonymous IEEE/ACM Conference]{IEEE/ACM Conference}{2022}{...} 

\usepackage{url}
\usepackage{algorithmic}
\usepackage{graphicx}
\usepackage{textcomp}
\usepackage{xcolor}
\usepackage{enumitem}
\usepackage{subcaption} 
\usepackage{array}
\usepackage{xcolor}
\usepackage{pifont}
\newcommand*\colourcheck[1]{%
  \expandafter\newcommand\csname #1check\endcsname{\textcolor{#1}{\ding{51}}}%
}
\colourcheck{blue}
\colourcheck{green}
\colourcheck{red}
\newcommand*\colourcross[1]{%
  \expandafter\newcommand\csname #1cross\endcsname{\textcolor{#1}{\ding{55}}}%
}
\colourcross{blue}
\colourcross{green}
\colourcross{red}

\newcolumntype{P}[1]{>{\centering\arraybackslash}p{#1}}

\usepackage[framemethod=TikZ]{mdframed}
\mdfsetup{
    roundcorner=5pt,
    backgroundcolor=gray!15,
    innertopmargin=5,
    innerleftmargin=5,
    innerbottommargin=5,
    innerrightmargin=5,
    skipabove=5,
    skipbelow=1,
    linewidth=1,
    nobreak=true
}

\begin{document}

\title{Effectiveness and Scalability of Fuzzing Techniques in CI/CD Pipelines} 


\author{Thijs Klooster}
\email{0000-0001-7151-1332}
\affiliation{
  \institution{TNO\\The Netherlands}
}
\author{Fatih Turkmen}
\email{0000-0002-6262-4869}
\affiliation{%
  \institution{University of Groningen\\The Netherlands}
}
\author{Gerben Broenink}
\email{0000-0002-3660-8179}
\affiliation{
  \institution{TNO\\The Netherlands}
}
\author{Ruben ten Hove}
\email{0000-0002-9137-9145}
\affiliation{
  \institution{TNO\\The Netherlands}
}
\author{Marcel Böhme}
\email{0000-0002-4470-1824}
\affiliation{
  \institution{MPI-SP\\Germany}
}

\begin{abstract}
Fuzzing has proven to be a fundamental technique to automated software testing but also a costly one. 
With the increased adoption of CI/CD practices in software development, a natural question to ask is `What are the best ways to integrate fuzzing into CI/CD pipelines considering the velocity in code changes and the automated delivery/deployment practices?'.
Indeed, a recent study by Böhme and Zhu \cite{Zhu2021RegressionFuzzing} shows that four in every five bugs have been introduced by recent code changes (i.e. regressions). 
In this paper, we take a close look at the integration of fuzzers to CI/CD pipelines from both automated software testing and continuous development angles. 
Firstly, we study an optimization opportunity to triage commits that do not require fuzzing and find, through experimental analysis, that the average fuzzing effort in CI/CD can be reduced by \textasciitilde 63\% in three of the nine libraries we analyzed (>40\% for six libraries).
Secondly, we investigate the impact of fuzzing campaign duration on the CI/CD process: A shorter fuzzing campaign such as 15 minutes (as opposed to the wisdom of 24 hours in the field) facilitates a faster pipeline and can still uncover important bugs, but may also reduce its capability to detect sophisticated bugs.  
Lastly, we discuss a prioritization strategy that automatically  assigns resources to fuzzing campaigns based on a set of predefined priority strategies. Our findings suggest that continuous fuzzing (as part of the automated testing in CI/CD) is indeed beneficial and there are many optimization opportunities to improve the effectiveness and scalability of fuzz testing.

\end{abstract}



\maketitle

\section{Introduction}\label{sec:1}

In the past decade, fuzzing has become one of the most successful automated vulnerability discovery techniques~\cite{liang2018fuzzing}. A fuzzer automatically generates inputs for a given Program Under Test (PUT) until the program crashes or a sanitizer terminates the PUT. A sanitizer is an automatic test oracle that crashes the PUT for inputs that expose security flaws, such as buffer overflows. Google uses 30k cores to continuously fuzz all Google products in \textsc{ClusterFuzz} \cite{google2017clusterfuzz}. Microsoft recently introduced fuzzing as a service in Project \textsc{OneFuzz}~\cite{Microsoft-onefuzzPlatform}. Google's \textsc{OSSFuzz} \cite{ossfuzz} continuously fuzzes more than 550 open source projects. In the five years since its launch, OSS-Fuzz has been continuously running and found more than 35k bugs. However, once a project has been sufficiently fuzzed, most new bugs found have been introduced by recent changes \cite{Zhu2021RegressionFuzzing}.

In this paper, we explore the challenges and opportunities of integrating fuzzing into the continuous integration / continuous deployment (CI/CD) pipeline where each pull request triggers an automatic build of the changed code which often runs automated regression tests. A \emph{pull request} (PR) is submitted by an author to integrate new code into the main branch of the code repository of the PUT. If the automated build fails, the PR authors and reviewers are informed accordingly. The integration of fuzzing into the CI/CD pipeline allows PR authors and reviewers to find bugs right when they are introduced and before they reach the main branch \cite{cifuzz,clusterfuzzlite,gitlab-cov-fuzz}.

Traditionally, fuzzing is considered resource and time intensive. In academia the recommended fuzzing campaign duration is at least \emph{24 hours}~\cite{evaluatingFuzzers18}. In the context of CI/CD, resources and time are costly and limited. \emph{10 minutes} is considered a reasonable build time (this includes testing)~\cite{fow,bjornw2018,Hilton2017Trade-offsFlexibility}. Furthermore, Github Actions currently charges 0.48 USD per hour \cite{githubCosts}, aborts automated builds after six (6) hours \cite{githubLimits}, and only limited resources are available, targeted at low intensity jobs.

Hence, we explore opportunities to save resources, e.g., by skipping fuzzing when we know for sure that no (new) bugs can be found as per the previous fuzzing sessions. In order to collect empirical evidence for this, we collected the most recent commits (ranging from 241 to 7847) of nine open-source libraries and analyzed them with a checksum-based method to help determine if fuzzing can be avoided after some of the commits. 
We also analyze the trade-off between fuzzing campaign duration and the number of bugs found for reasonable values of campaign duration. Our hypothesis is if fuzzing a non-fuzzed target can uncover many vulnerabilities with relatively few resources~\cite{Bohme2020}, shorter but continuous fuzzing suitable for CI/CD pipelines should still be beneficial. 

For our evaluation, we chose the \texttt{Magma} state-of-the-art fuzzing benchmark, and thus the open source libraries it supports, because it provides a ground truth~\cite{hazimeh2020magma}. Our first set of experiments revealed that in one third of the libraries we analyzed, many fuzzing sessions (\textasciitilde 63\% of all sessions) can be skipped even if the PUT has changed. This is more than 40\% for six libraries. We also measured the bug finding effectiveness for nine reasonable values of fuzzing campaign durations, ranging from 5 minutes to 8 hours. Our foremost observation from this experiment is that running a fuzzing campaign of 15 minutes (slightly longer than the recommended 10 minutes of build and test time in CI/CD) can be effective. Indeed, for half of the libraries at least one bug is reached, triggered and detected by fuzzing. Increasing the duration by 5 minutes overall results in higher bug detection rates.

Our work provides empirical evidence that:
\begin{itemize}
    \item Considerable fuzzing effort can be avoided, thereby saving computational resources, when source code changes (i.e., commits) of the PUT are carefully scrutinized.
    \item Continuous but shorter fuzzing (e.g., 15 minutes) campaigns can still uncover important bugs.
\end{itemize}  
One of the most practical implications of these results is that fuzzing campaigns in CI/CD pipelines running on moderate infrastructures can significantly benefit from fuzzing, when applying certain optimizations and trade-offs. Finally, our work opens new interesting directions of research for optimizing not only fuzzing methods themselves but also how fuzzers can be used more efficiently.

\section{Background and Motivation}\label{sec:2}
In this section, we summarize fuzzing as an automated testing technique and motivate the need for its integration to CI/CD pipelines. The section ends with a set of concrete research questions we aim to answer in our work. 

\subsection{Fuzzing in CI/CD}
\emph{Fuzzing} is a dynamic testing technique that, starting from one or more seeds, executes a PUT with repeatedly mutated inputs until some “interesting” behavior is observed~\cite{evaluatingFuzzers18}. A fuzzer, implementing a fuzzing technique, records the observed program behaviour and saves the inputs that lead to the observed behaviour for future use. The fuzzing process generally stops in one of the following cases:
\begin{itemize}
\item the user intervenes, 
\item a predetermined timeout is reached,
\item a certain bug type is found. 
\end{itemize}

Fuzzers are classified according to their knowledge of the PUT in terms of input format and program structure (i.e., white-box, grey-box and black-box), and the strategies they employ (e.g., mutational fuzzers) in generating inputs. A \emph{black-box fuzzer} assumes no knowledge of the PUT, a \emph{grey-box fuzzer} assumes partial knowledge and often leverages code instrumentation to glean that knowledge, and a \emph{white-box fuzzer} employs sophisticated analysis techniques (e.g., symbolic execution).
In order for a fuzzer to test the robustness of some functionality within a software library, a \emph{fuzzing harness} needs to be written, which acts as the main entry point for the fuzzer to reach the specific functionality.
A single library can have many fuzzing harnesses, each providing access to different kinds of functionality of the software.
In the remainder of this work, we use the term \emph{fuzz target} to indicate a compiled fuzzing harness (i.e. the executable file resulting from the compilation of the harness).

A \emph{CI/CD pipeline} refers to the largely automated process of integrating committed changes to the code, testing and then moving the code from the commit stage to the production stage~\cite{paule2018securing}.
Frequent but small (merge) commits are encouraged~\cite{fow, CIBook2007}. Indeed, a large scale analysis conducted by Zhao et al.~\cite{Zhao17} determined around 21 (median) commits are made per day for 575 open source projects. Performing automated testing on each code change enables earlier feedback~\cite{fow} on bugs since failed stages and errors can be traced to sources (and the developers) more quickly, thanks to the limited scope~\cite{humble2010continuous}.

However, fuzzing is often a resource-intensive process and starting a long-running fuzzing campaign after each commit may not scale. Since only some commits cause code changes that require fuzzing, methods to identify such commits are needed to integrate fuzzers into CI/CD pipelines.

Moreover, determining a good timeout for fuzzing, also called fuzzing campaign duration, is crucial for effective testing. While there are different proposals, a commonly accepted fuzzing campaign duration is 24 hours~\cite{evaluatingFuzzers18}. However, in a CI/CD setting, we are limited in terms of the available resources~\cite{memon2017taming} and running an automated build for 24 hours on each commit is unacceptable. Indeed, in a CI/CD setting the testing time that is considered "reasonable" is much shorter. For instance, Fowler~\cite{fow} proposes 10 minutes as a guideline. Thus, striking a balance between fuzzer effectiveness and testing duration in the CI/CD pipelines is of practical concern.

The idea of continuous fuzzing (in the CI/CD sense) is not new. For instance, Google offers \texttt{CIFuzz}~\cite{GoogleIntegration} as part of \texttt{OSS-Fuzz} that runs fuzzers for 10 minutes (max 6 hours) over 30 day old/public regressions of (selected) open source projects. Similarly, GitLab has coverage-guided fuzzing (and Web API fuzzing) integrated to its CI/CD offerings that can be run either 10 minutes or 60 minutes~\cite{gitlab-cov-fuzz} in its own pipeline stage. However, there is no empirical study investigating this trade-off between bug finding and campaign length that can inform this decision.

\subsection{Existing Optimizations of Fuzzing}\label{sec:existing_optimizations}
In order to improve fuzz testing, many optimizations have been proposed that either aim to improve general fuzzer efficiency, or the effectiveness of (continuous) fuzzing practices.
Since this work concerns a study on the effectiveness of fuzzers in the context of CI/CD, we focus mainly on existing approaches that impact the fuzzing performance within this context, and therefore do not go into detail on general optimizations. A combination of these existing approaches forms the building blocks of the optimizations proposed in our work.



\paragraph{Regression fuzzing}
Standard fuzzing sessions (not in the context of CI/CD) build all fuzz targets and start fuzzing each of them typically for longer periods of time (e.g. 24 hours).
In contrast, regression fuzzing  runs for shorter periods of time (e.g. 10 minutes) in order to let the continuous integration pipeline finish within a reasonable amount of time~\cite{gitlab-cov-fuzz}.
This implies that there are far less computational resources available for the actual fuzzing, yet this type of fuzzing focuses more on finding bugs that are regressions in the code (bugs related to a feature that has worked before stop working after the recent changes), which still makes regression fuzzing quite valuable~\cite{Zhu2021RegressionFuzzing}.
From this perspective, a relatively short fuzzing campaign duration can still prove to be beneficial.
In practice, regression fuzzing is facilitated through corpus sharing and minimization~\cite{google2017clusterfuzz} which are summarized below.

\paragraph{Corpus sharing}
Corpus sharing lets new fuzzing sessions build upon the progress of past fuzzing sessions.
This is achieved by adding all of the newly found interesting inputs to the existing corpus, and in turn using the augmented corpus as the seed corpus for the new fuzzing session.
Using this approach avoids the repeated coverage of the parts of software that were already covered by previous fuzzing sessions.
These computational resources conserved in this way can now be utilized to push the boundaries of coverage further into the software under test.
Considering this approach generates many new test cases, the corpus has to be minimized periodically to prevent exponential growth in size~\cite{herrera2019corpus}. 

\paragraph{Corpus minimization}
Corpus minimization is a process that takes an existing corpus and its corresponding fuzz target, and then filters out the inputs within the corpus that are superfluous (that do not reach unique parts of the fuzz target).
Using this approach, only the most important inputs remain, such that the corpus is still as effective, yet smaller in size.
In general, it is the case that the smaller the corpus, the higher its performance will be (when comparing the before and after of minimization applied to a corpus).

\paragraph{Ensemble fuzzing}
There is also an approach called ensemble fuzzing, which is a form of fuzzing where multiple fuzzer processes work together on the same fuzz target.
A fuzzing session can consist of a single process that fuzzes a single target, but it is also possible to have several fuzzer processes that work on the same target with the same shared corpus, such that each process can build upon the progress of its peers.
This technique can speed up the effort and reach certain parts of the software under test more quickly.
While the processes that work together can be based on one and the same fuzzer, it is also possible for different kinds of fuzzers to collaborate sharing the corpus and thus sharing the progress~\cite{chen2019enfuzz,osterlund2021collabfuzz}.
The idea behind this technique is that each of the fuzzers is more capable in solving and reaching one part of the software, while others may be more specialized in reaching other parts, and so a collaboration between these fuzzers makes the overall process more efficient.

\paragraph{Sanitizers}
There is a type of tool known as sanitizer, which can aid fuzzers with their fault-detection capabilities.
Sanitizers basically consist of a large set of fault-detecting code snippets that get injected into the fuzz targets at compilation time.
For example, ASan (AddressSanitizer~\cite{serebryany2012addresssanitizer}) is a tool that can detect errors such as out-of-bounds and use-after-free.
It is able to do this by instrumenting the code that is being compiled, such that the tool can detect specific faults if they occur while the fuzzing effort runs.
Seeing as fuzzers detect when a fault has occurred through a crash of the target program, there are faults that a standalone fuzzer cannot detect reliably.
Some faults can occur without crashing the target program, and so while the fault has been triggered, this is not being detected by the fuzzer.
This advocates the importance of employing these tools at least once in a while, such that these bugs do not go unnoticed forever.
The downside of using sanitizers is that they slow down the fuzzer process (in the case of ASan a slowdown of 2x can be expected~\cite{asan}).
There are other sanitizers that can be used as well, each of them specializing in the detection of certain faults, but these too come at the cost of a slowdown of the process.

\subsection{Research Questions}


In our work we are interested in the dimensions that will have the most impact on the performance of the fuzzing effort in CI/CD pipelines. More specifically we want to answer the following questions:
\begin{itemize}
    \item[\textbf{RQ.1}] \textbf{How to fuzz?} 
    How can we exploit different fuzzers in a \emph{continuous fuzzing} setting?
    \item[\textbf{RQ.2}] \textbf{What to fuzz?} Can we avoid some of the fuzzing effort by carefully selecting fuzz targets?
    \item[\textbf{RQ.3}] \textbf{How long to fuzz?} What is a reasonable fuzzing campaign duration that is compatible with CI/CD testing timelines but is still effective in finding bugs? 
\end{itemize}

\section{Trade-offs in Continuous Fuzzing}\label{sec:3}
In the following, we explore several trade-offs when using fuzzing in continuous software development practices (i.e., in CI/CD).
One trade-off is related to the limited usage of computational resources while employing continuous fuzzing techniques extensively. It is aimed at analyzing and possibly avoiding the fuzzing campaigns for harnesses that remain semantically unchanged by the current commit.
Another trade-off is about finding a  reasonable duration for fuzzing campaigns where striking a balance between build time and fuzzer effectiveness is the main concern. Given the exponential cost of vulnerability discovery in fuzzing \cite{Bohme2020}, it should be possible to find a reasonable number of bugs at a reasonably low cost.

\subsection{Fuzzing Smarter, Not Harder}
Within the program under test, there may be one or more fuzzing harnesses, and each of those harnesses provides a different entry point into the software for the fuzzer. 
Fuzzing the compiled harnesses is an effort which is highly parallelizable, and so in order to maximize performance, each of these targets can be fuzzed on their own CPU-core.
This approach works quite well for standard fuzzing sessions which are executed every now and then, and have a duration upwards of 24 hours.
However, looking at fuzzing in the context of continuous integration, this approach does not scale well.
This is because in these pipelines, there can be many commits, on possibly multiple branches of the repository.

If we want to launch the fuzzing effort for every new commit that is pushed to the repository, this demands a large amount of computational resources.
Depending on the frequency of the commits and the time allowed for the pipeline to run, this can very quickly use up all of the resources that are available for fuzzing, and in doing so create a large queue of fuzzing sessions that have yet to start.
Such a queue can be dealt with in several ways, one of which would be to simply keep track of them and start the pipeline for each of them one after the other.
Interrupting the currently running fuzzing process when a new commit is pushed would be a different technique that can be applied in this situation.
Another approach can be to only start processing the most recent commit at the time of a previous pipeline completion (as opposed to processing all commits in between as well), seeing as the commits in between would already be outdated.
But in the case where the fuzzing process selectively fuzzes only targets affected by a commit, no commits should be skipped (otherwise the respective code changes will never be fuzzed).
Typically, a CI/CD pipeline makes use of commonly available and cheap compute resources in order to execute the different stages of the pipeline.
Although these resources can also be used for fuzzing, it is often the case that these resources are quite limited and therefore negatively impact the performance of fuzzing campaigns.
While these commonly available resources are sufficient for static application security testing, the dynamic nature of fuzzing requires more computational power.

In an effort to prevent the consumption of excessive amounts of resources, we can employ techniques to decrease the demand of resources that comes with every new commit that is pushed.
Since per-commit fuzzing sessions use regression fuzzing instead of standard fuzzing sessions, this facilitates the possibility of skipping the fuzzing campaigns for unaffected targets.
The idea is to optimize the fuzzing stage of the pipeline in such a way that it will only fuzz the targets that are actually affected by the code changes from the commit that was pushed.
In the ideal case, this would mean tracing the affected lines of code to a specific fuzzing harness, compiling only that harness, and then start fuzzing it (instead of fuzzing all harnesses).
The fuzzing stage will ideally be very focused as well, steering the fuzzer towards the affected parts of the fuzz target.
Knowing which of the fuzzing harnesses are - directly or indirectly - affected by the code changes of a certain commit is not trivial.

We propose to make a selection of which targets to fuzz based on whether a target differs from the same target of the preceding, already fuzzed commit.
If both versions of the compiled target are syntactically equivalent, we know \emph{for sure} that their behavior is equivalent, as well. No new bugs could have been introduced in this commit. Since the previous version was already fuzzed, fuzzing this version is redundant.
If they are different, the functionality might have changed and it is worthwhile to fuzz this commit.
To be fair, this approach may not fully reflect the ground truth of affected fuzz targets, seeing as only changes that are directly related to the fuzz targets are taken into consideration for this simple approach.
Any indirect changes that affect the functionality of the fuzz target can only be accounted for when implementing a more elaborate technique for target selection.
Nonetheless, our aim is to investigate whether using a simple approach can result in a significant reduction of the number of fuzz targets that require resources. We leave more elaborate techniques for redundancy optimization as future work.



\subsection{Fuzzing Duration versus Effectiveness}
\label{sec:fuzzdur}
We are interested in the degree to which the bug finding ability of a fuzzer is affected when the duration of the fuzzing campaigns is relatively short.
The longer a campaign lasts, the more work a fuzzer is able to do. Standard fuzzing sessions can last 24 or even 72 hours.
However, within the CI/CD pipeline we are time-constrained.
After they made the commit, developers do not want to wait on long-running fuzzing campaigns, but obtain results within a reasonable amount of time.
Additionally, widely adopted DevOps tools like GitHub Actions even terminate jobs that exceed the 6 hour mark~\cite{githubactionslimit}.
Fowler \cite{fow} states that a build time of ten minutes would be a good guideline, confirmed by Hilton et al. \cite{Hilton2017Trade-offsFlexibility} stating that the maximum acceptable build time is ten minutes, based on the most common answer from their survey participants.
From the perspective of developers, this suggests that fuzzing campaigns should not last longer than ten minutes.
Because of the conflicting interests between developers wanting shorter builds and fuzzers wanting longer builds, we aim to gain more insight into the effects of the choice of fuzzing campaign duration on the effectiveness of the fuzzing process.

As is nicely illustrated by Hilton et al. \cite{Hilton2017Trade-offsFlexibility}, the duration of a build could be described by \textit{teatime}, \textit{lunchtime}, or \textit{bedtime}. 
Teatime represents the shorter builds (10 minutes), lunchtime represents the medium builds (1 hour), and bedtime represents the longer builds (8 hours).
This suggests there could be several times of day at which a developer may not mind having the longer build times, if this means allowing for the fuzzing campaign to make more progress.
Teatime may be the default setting for the fuzz duration, while enabling lunchtime or bedtime for a certain commit signifies to the pipeline that additional fuzzing time is allowed for that specific build.
In practice, this can be implemented as a setting which can be enabled manually, but the fuzzing campaign duration may even be determined automatically based on some priority that could be generated for each commit. 
In our work, we are investigating the bug finding ability of continuous fuzzing for reasonable values of campaign duration.

\section{Benchmarks}\label{sec:4}
In order to investigate the effectiveness of fuzzers while varying the fuzzing campaign duration in the CI/CD setting, we need a fuzzing platform that enables the evaluation of multiple fuzzers and supports good quality benchmark suites (i.e. target programs). While there exist several such platforms, we chose to employ \texttt{Magma}~\cite{hazimeh2020magma} in our experiment due to the following reasons:
\begin{itemize}
    \item \texttt{Magma} is a state-of-the-art fuzzing benchmark with an emphasis on providing the \textit{ground-truth} for bugs and their precise location in the PUT. This is achieved by reverting the fixes for known, real-world bugs which have existed in their respective software repositories.
    \item The collection of target programs consists of nine distinct open source software projects that are widely used.
    \item The functionality of target programs in the benchmark covers a wide range of applications with varying complexities.
    \item The benchmark provides instrumented targets which make a distinction whether a bug was \textit{reached}, \textit{triggered}, or \textit{detected} to enable more accurate evaluation of fuzzer effectiveness.
    \item \texttt{Magma} supports many different fuzzers out-of-the-box, including \texttt{AFL++}, \texttt{libFuzzer}, and \texttt{Honggfuzz}, which are used in Google's \texttt{OSS-Fuzz} platform as well~\cite{ossfuzz}.
    \item The benchmark is highly configurable and its setup is easily adaptable for running custom fuzzing campaign evaluations.
\end{itemize}

\begin{table*}
    \centering
    \caption{CI/CD details and \texttt{Magma} parameters of the nine libraries used in the experiments (as of May 4th, 2022).}
    \label{tab:Magma-libs}
    \begin{tabular}{|l||r|r|r|c||r|r|}
         \hline
         {Library} & {Commits} & {Branches} &  {Repository size (MB)} & {Language} & {\texttt{Magma} fuzz targets} & {\texttt{Magma} bugs} \\
         \hline \hline
         \texttt{libsndfile} & 3083   & 23       & 36  & C   & 1 & 18 \\         
         \texttt{libtiff}    & 3953   & 4        & 17  & C   & 2 & 14 \\
         \texttt{libpng}     & 4098   & 8        & 48  & C   & 1 & 7  \\         
         \texttt{libxml2}    & 5353   & 12       & 52  & C   & 2 & 17 \\
         \texttt{lua}        & 5420   & 5        & 13  & C   & 1 & 4  \\
         \texttt{poppler}    & 7206   & 20       & 24  & C++ & 3 & 22 \\
         \texttt{sqlite3}    & 26610  & 1128     & 146 & C   & 1 & 20 \\
         \texttt{openssl}    & 31163  & 21       & 662 & C   & 5 & 20 \\
         \texttt{php}        & 128239 & 407      & 615 & C   & 4 & 16 \\
         \hline
    \end{tabular}
\end{table*}

Since we decided to employ the \texttt{Magma} benchmarking tools in our experiments, this provides us with a solid set of software libraries to fuzz. 
All of these libraries are listed in Table \ref{tab:Magma-libs} which also includes various CI/CD aspects of their online repositories. 
For each library, the number of fuzzing targets that can be used for benchmarking varies.
Since the \texttt{Magma} benchmark is created by reintroducing known bugs that existed in the library in the past, the number of bugs  per library varies as well.
These software libraries are still receiving commits to this day, which means they are still actively developed, and fuzzing them has real-world and practical benefits.

While there are several metrics that are being used for fuzzer evaluation in practice, \textit{the number of crashes} is arguably the most used. Klees et al.~\cite{evaluatingFuzzers18} suggest that the number of crashes should ideally reflect the number of actual bugs.
In conjunction with the reasons mentioned earlier, this makes \texttt{Magma} a sublime candidate for our evaluation of continuous fuzzing.
However, \texttt{Magma} is intended to be used for standard fuzzer evaluation, not specifically in the context of continuous fuzzing.
Therefore, we have extended the benchmark with the optimizations summarized in Section~\ref{sec:existing_optimizations}. In the following section, we provide more details about the integration of these techniques to our Magma-based benchmarking setup. 

\subsection{Setup}
The \texttt{Magma} benchmark provides a platform for evaluating fuzzers by including several metrics, multiple libraries, and many known bugs.
This platform orchestrates the individual fuzzing campaigns, monitors the findings, and outputs the results.
The platform also allows for the specification of parameters of the fuzzing campaigns, e.g. which fuzzers to use, which fuzz targets to use, and the amount of repeated trials.
Since the platform is based on Docker, it was fairly easy to adapt it to our needs for evaluating the effectiveness of continuous fuzzing.
In our setup, we employ multiple fuzzers in a continuous manner (i.e., continuous ensemble fuzzing) however the duration of fuzzing campaigns is significantly shorter in comparison to traditional ensemble fuzzing. 
Continuous ensemble fuzzing can be very beneficial when integrated into CI/CD pipelines, seeing as \texttt{OSS-Fuzz} has managed to find 40k bugs in five years~\cite{ossfuzzbugs} while employing such an ensemble fuzzing approach.
The utilization of multiple fuzzers results in an increase in resource usage, though the empirical evidence provided by \texttt{OSS-Fuzz} suggests that it is an effective way of fuzzing in continuous practices.
Therefore, this will be our starting point from which we aim to reduce resource usage based on fuzz targets that have not been affected by code changes.
Whenever there are targets that do not have to be fuzzed, this actually results in 3 less CPU cores needed per target (when using an ensemble setup of the 3 fuzzers that \texttt{OSS-Fuzz} uses).

\begin{figure*}
\centering
\includegraphics[width=0.8\linewidth]{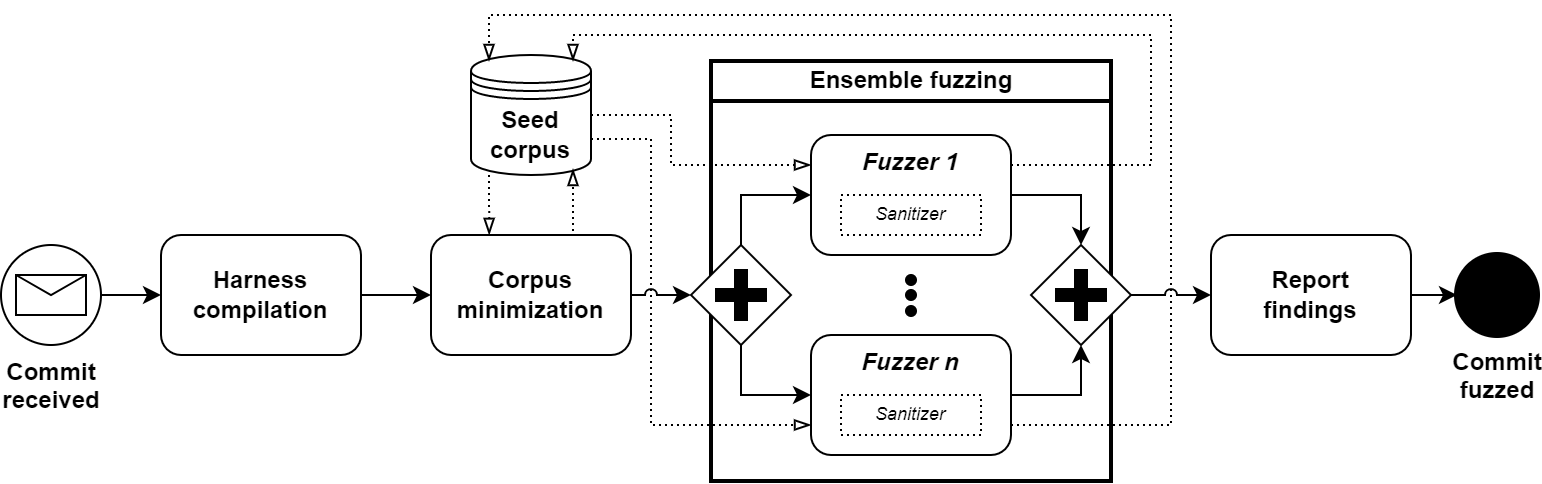}
\caption{General design of ensemble fuzzing in the context of continuous integration.}
\label{ensemble}
\end{figure*}

\autoref{ensemble} shows the overall design of ensemble fuzzing in the context of continuous integration.
Upon receiving a new commit, the harnesses will be compiled into targets, which in turn will be used to minimize the corpus.
The minimized corpus is used by every fuzzer in the setup, where every fuzzer contributes towards an updated corpus by submitting newly found interesting inputs to that corpus.
This updated corpus will be re-used in future commits, effectively allowing fuzzers to collaborate on each fuzz target.
Corpus sharing is implemented this way in \texttt{ClusterFuzzLite} and in GitLab's coverage-guided fuzzing as well~\cite{clusterfuzzlite,gitlab-cov-fuzz}.
Lastly, when the predetermined timeout has been reached, the fuzzers will stop and report their findings.

The Magma benchmark is based on real-world bugs that existed in the software before, and provides a specific version for each of the programs used, where all of the bug-fixes are reverted to this version.
This implies that we cannot use real-world commits, otherwise we would lose the advantage of having the ground-truth \texttt{Magma} bugs. Hence, we simulate commits by repeating fuzzing campaigns multiple times.
Ensemble fuzzing is not supported out-of-the-box in the Magma benchmark, since the benchmark is intended for evaluating the performance of individual fuzzers.
In standard fuzzing sessions that take up to 24 hours or even longer, ensemble fuzzing requires fuzzers to share the same working directory, such that their progress can be shared throughout the whole of the campaign.
In continuous fuzzing however, where the average per-commit campaign duration would be much shorter, e.g., 15 minutes, it suffices to synchronize the fuzzers after every commit, combining the corpora and minimizing the result to obtain the new seed corpus for the next commit that will get pushed.
The extension of the \texttt{Magma} benchmark with these techniques makes for a great platform to base our continuous fuzzing evaluation on.

\begin{mdframed}
In order to address \textbf{RQ.1}, a continuous fuzzing setup where different fuzzers can collaborate on the same targets effectively can be designed by adapting \textit{ensemble fuzzing} for continuous integration, \textit{sharing} and \textit{minimizing} the corpus \textit{between each commit}.
\end{mdframed}

\section{Experiments}\label{sec:5}

We conducted an extensive set of experiments in order to seek answers to the questions \textbf{RQ.2} and \textbf{RQ.3}. In this section, we elaborate upon the implementation of the experiments and present the results.

The source code of the extended \texttt{Magma} benchmark as well as the source code used for the experiments and the plots that have been generated from the experiments is open-source and can be found online\footnote{\url{https://github.com/kloostert/CICDFuzzBench}}.

\subsection{Fuzz Target Selection}
In an effort to optimize the resource usage in continuous fuzzing, the first experiment aims to find out what proportion of the fuzzing campaigns should be started by default. In other words, we are interested in understanding if there are any commits that should not trigger a fuzzing campaign due to the code changes that do not affect certain fuzz targets.

One option of determining whether a fuzz target is different from that of the preceding commit, is by calculating the checksum of both targets.
Calculating a checksum can be done over the source code of the target although it is still not trivial to trace which lines of code affect which target given the checksums.
Instead, calculating the checksum of the fuzz target itself (being an executable file) would present even a simpler solution and reveal whether the target is different between the respective commits.
Before the checksum can be calculated, the fuzzing harnesses have to be compiled into targets.
This does require some time and resources, in contrast to what we would be able to do it without compiling the harnesses.
Fortunately, the compilation of source code in a CI/CD pipeline is oftentimes already taken care of in the building stage, and so this compilation will not result in any additional overhead during the fuzzing stage.
After compilation, checksums will be calculated for all of the targets, and they will be compared to the checksums from the preceding commit.
\emph{For targets that have identical checksums, we can decide to not start their corresponding fuzzing sessions}, saving computational resources.

Another option is to check whether there is a change in code coverage. For instance, Google's \texttt{CIFuzz}~\cite{GoogleIntegration} also provides the option to employ target selection. Yet, this option is available only to projects that support \texttt{OSS-Fuzz}'s code coverage, which makes it more complex to set up properly.
Using checksums directly on fuzz targets enables software projects for which \texttt{OSS-Fuzz}'s code coverage is not available to take advantage of the fuzz target selection.

Table \ref{tab:targsel} lists the number of fuzzing harnesses that were processed during the target selection experiment, the number of commits that were processed for each library, and the proportion of the processed commits over the total amount of commits in the respective repositories.
Since \texttt{Magma} is used throughout our experiments, we restrict ourselves to the nine  software libraries that are included in \texttt{Magma} as the data set for the target selection experiment.
Instead of using \texttt{Magma}'s version of the libraries that contain the bugs, this experiment focuses on the live repository, and so a selected number of the most recent commits for all of the nine libraries are processed.
On a side, we observed that the number of fuzzing harnesses available on the live repositories differ from those available in the \texttt{Magma} benchmark for certain libraries.

When the experiment starts, the most recent version of the software is taken, and the processing will continue backwards in time from there.
For every commit, checksums for all of the available fuzzing harnesses will be obtained by first compiling the harnesses into targets, and consecutively calculating their checksum.
The same will be done for the preceding commit, after which the checksums will be compared for each target, resulting in a proportion of unchanged fuzz targets for a single commit.
These steps will be repeated until either the repository changed in a way that automated fuzz target compilation was no longer possible, or the fuzzing harnesses themselves were no longer part of the repository (the commit that introduced the fuzzing harnesses was reached).
When this process terminates, we have obtained the overall proportion of fuzz targets that remained unaffected by commits on average.

\begin{table}
    \caption{Target selection experiment details: the number of harnesses and commits which were processed (along with the proportion of all commits in the repository), and the proportion of identical fuzz targets that was found. ( * Two of the libraries include versioning info in the executable fuzz targets, which interferes with the checksum approach.)}
    \label{tab:targsel}
    \centering
    \begin{tabular}{|l||r|r|r|r|r|}
         \hline
         {Library} & {Harnesses} & {Commits} & {Identical} \\
         & processed & processed & targets\\
         \hline \hline
         \texttt{libsndfile} & 1   & 241 (8\%)   & 64\% \\         
         \texttt{libtiff}    & 2   & 801 (21\%)   & 53\% \\
         \texttt{libpng}     & 1   & 1158 (28\%)  & 41\% \\
         \texttt{lua}        & 1   & 2285 (42\%)  & 20\% \\
         \texttt{poppler}    & 2   & 1919 (27\%)  & 44\% \\
         \texttt{openssl}    & 12  & 7847 (26\%)  & 63\% \\
         \texttt{php}        & 9   & 7821 (6\%)  & 64\% \\
                             &     &              & \textit{Weighted mean:} 55\% \\
         \hline \hline         
         \texttt{libxml2}*   & 1   & 625 (12\%)   & 00\% \\
         \texttt{sqlite3}*   & 1   & 483 (2\%)   & 00\% \\
         \hline
    \end{tabular}
\end{table}

\textbf{Results}. As mentioned in Table \ref{tab:targsel}, the experiment achieved varying proportions of processed commits among the nine libraries, ranging between 6 and 42 percent of all commits.
During the experiment, it also became clear that two of the nine libraries (\texttt{libxml2} and \texttt{sqlite3}) never produced identical fuzz targets.
Further investigation showed that this was due to the inclusion of versioning information (compilation timestamp or \texttt{Git} revision) in the actual executable fuzz targets, interfering with the checksum approach for target selection.
In order to alleviate this problem, either a more sophisticated form of target selection has to be applied, or the build process of these fuzz targets should be adapted in such a way that no versioning information makes it into the compiled targets.
Either way, since the proportion of identical fuzz targets could not be measured properly for these two libraries, \texttt{libxml2} and \texttt{sqlite3} are not taken into consideration towards the final results of this experiment.

For the remaining seven libraries, the proportion of identical fuzz targets ranges from 20 to 64 percent, where the weighted arithmetic mean lies at 55 percent (as shown in Table \ref{tab:targsel}).
\begin{mdframed}
These results suggest that even employing a \textit{simple} technique for selecting the subset of fuzz targets which should be fuzzed, can result in saving \textit{more than half} of the computational resources needed for the continuous fuzzing effort.
\end{mdframed}
Another observation is that the scope of the fuzzing harnesses greatly impacts the proportion of identical targets, since harnesses that are aimed specifically at one single function of the target library relatively do not receive many changes, while harnesses that cover a library more broadly (like \texttt{lua}) receive relatively more changes.

\begin{figure*}
\centering
\includegraphics[width=0.8\linewidth]{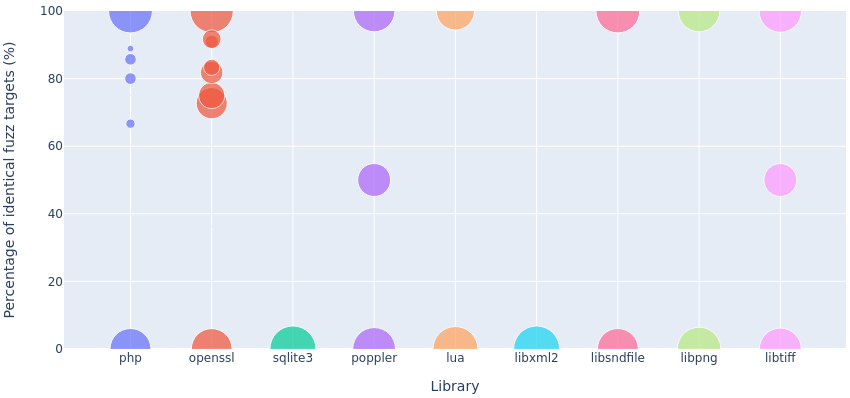}
\caption{The distribution of the percentage of identical fuzz targets per library for all processed commits. The marker size represents the logarithmically scaled distribution per library. The libraries \texttt{libxml2} and \texttt{sqlite3} have no identical fuzz targets.}
\label{bubble}
\end{figure*}

Figure \ref{bubble} shows more details across fuzz targets and commits per project. Exploring the results more in-depth, the experiment reveals that the distribution of identical fuzz targets per commit varies among the seven libraries greatly.
We find that for most of the commits, either none of the fuzz targets were affected (the complete fuzzing stage of the pipeline can be skipped), or all of the targets were affected (all fuzz targets are fuzzed simultaneously).
Three of the libraries (\texttt{libsndfile}, \texttt{libpng}, and \texttt{lua}) do not show any commits with a proportion of identical commits somewhere between 0 and 100 percent, since only a single fuzz target was processed for these libraries.
Two other libraries (\texttt{libtiff} and \texttt{poppler}) do have commits where only one of the two fuzz targets was affected.
The libraries with the largest number of fuzz targets (\texttt{openssl} and \texttt{php}) show that most commits fall into the categories of either no targets were affected or all of them were, however we also observe a significant number of commits where the code changes affected only a small portion of the fuzzing harnesses.
Target selection in a CI/CD pipeline would be beneficial in those cases as well, since the number of needed CPU cores scales with the number of affected fuzz targets.
\begin{mdframed}
In order to address \textbf{RQ.2}, we found that it suffices to \textit{only fuzz the targets for which we know the code has changed} by the latest commit. Compared to fuzzing all of the targets every time, this would \textit{save up to 64 percent} in terms of fuzzing campaigns, and therefore in the computational resources required for fuzzing. 
\end{mdframed}

\subsection{Fuzzing Campaign Duration}
In an effort to settle the dispute between CI/CD pipeline durations and their effectiveness, the second experiment aims to find the optimal campaign duration that strikes a balance between quick builds and fuzzer effectiveness.
For this experiment, we chose to include three of the most popular fuzzers that are successfully being used in modern fuzzing platforms like \texttt{OSS-Fuzz}~\cite{Serebryany2017}: \texttt{AFL++}~\cite{aflplusplus}, \texttt{libFuzzer}~\cite{libFuzzer}, and \texttt{Honggfuzz}~\cite{hongfuzz}.
The choice of software libraries to benchmark the fuzzing campaigns is elaborated upon in Section \ref{sec:4}, while their details are presented in Table \ref{tab:Magma-libs}.
In order to deal with the stochastic nature of fuzzers, we repeat every fuzzing campaign 10 times to reduce the effect of outliers due to randomness.
These repetitions of one and the same fuzz duration are used to simulate 10 commits as well.
For every fuzz duration, these successive runs will build a corpus along the way, as would be the case with real commits.
Even though there are no actual code changes in these simulated commits, this does not affect the results since the effectiveness of the fuzzing campaigns is not linked to code changes (as would be the case with directed fuzzing, where a fuzzer is specifically directed to these changes).

As described in Section \ref{sec:fuzzdur}, we want to investigate the impact on the effectiveness of continuous fuzzing when significantly reducing the duration of fuzzing campaigns.
Towards this effort, we arrived at a set of campaign durations that would be of interest for the experiment, ranging from 5 minutes to 8 hours.
The experiment was conducted on a Linux virtual machine with 8 vCPUS running at 2.1 Ghz, 8 GB of RAM, and 80 GB of disk space.
\begin{mdframed}
During our fuzzing campaign duration experiment, we conducted \textit{over 1 CPU-year worth} of continuous fuzzing campaigns.
\end{mdframed}

The experimental setup uses corpus sharing and minimization between successive repetitions of the fuzzing campaigns, in order to simulate 10 commits being processed in a continuous fuzzing setting that would use the same techniques.
Logically, this is a realistic assumption, seeing as fuzzing campaigns that do not continue where the ones from previous commits left off, are not expected to be very effective.
Simulating commits instead of using ones from the live repositories is done because (by using \texttt{Magma}) we know exactly where the bugs are and when they are triggered by the fuzzers.
This is not the case for live commits, where ideally no bugs should be present at all.
For the experiment, we also start off with a given set of initial seed inputs for the fuzz targets, otherwise it would take quite a few commits to obtain basic coverage of the fuzz targets.
Moreover, starting from a given seed corpus is realistic as well, as such corpora will be generated through longer-running fuzzing campaigns anyway, while initializing a CI/CD pipeline to include continuous fuzzing.
Last but not least, \texttt{AFL++}, \texttt{libFuzzer}, and \texttt{Honggfuzz} will be working together in the experiment (through corpus sharing), such that they can build upon the progress of their peers, which will be the case in a real continuous fuzzing setup as well.

\textbf{Results}. While running the experiment, one observation was that for \texttt{lua}, no bugs were found within any of the time budgets.
This is due to the fact that there are only 4 bugs included in \texttt{Magma}'s version of \texttt{lua}, where other libraries can have upwards of 20 different bugs available (see Table \ref{tab:Magma-libs}).
Apparently these 4 bugs are quite complex ones, in the sense that finding them through fuzzing requires more time than we allocated to the campaigns.
Another observation was that, looking at the number of bugs \textit{detected}, for some libraries there is no visible difference between 15 minute and 8 hour campaign durations, while for other libraries there is.
It seems like the libraries with less commits benefit more from having 8 hour fuzzing campaigns than the libraries with more commits.
\begin{mdframed}
This suggests that while employing \textit{per-commit} fuzzing campaigns of \textit{15 minutes} in general strikes a solid balance between the desires of developers and the effectiveness of fuzzing, it is worthwhile to add in \textit{8 hour} fuzzing campaigns on snapshots of the repository \textit{every now and again}.
\end{mdframed}
This would ideally take the form of \textit{bedtime} fuzzing as mentioned in Section \ref{sec:fuzzdur}, where during the night the current version of the library is fuzzed, further fueling the short fuzzing campaigns to come during the next day.
This way, regressions can be discovered very quickly, while the more difficult-to-find bugs are caught during the night.

\begin{figure*}
     \centering
     
     \begin{subfigure}[b]{\textwidth}
         \centering
         \caption{The mean number of \textit{reached} \texttt{Magma} bugs.}
         \includegraphics[width=\textwidth]{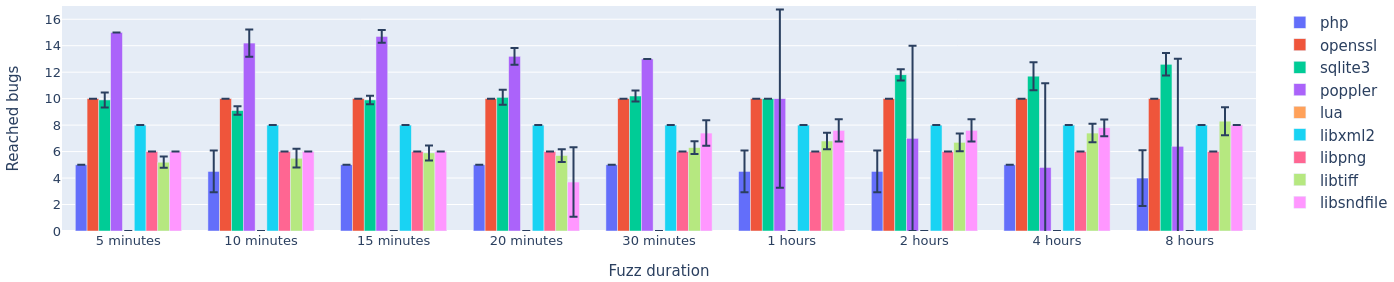}
         \label{reached}
     \end{subfigure}
     
     \begin{subfigure}[b]{\textwidth}
         \centering
         \caption{The mean number of \textit{triggered} \texttt{Magma} bugs.}
         \includegraphics[width=\textwidth]{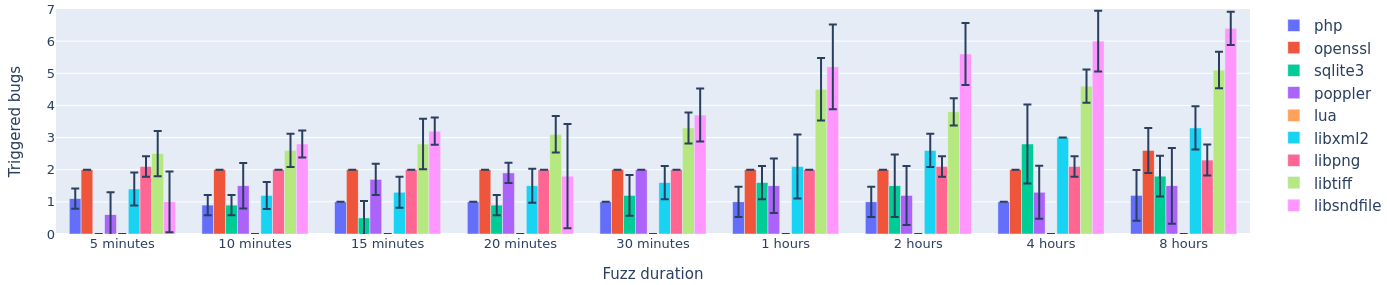}
         \label{triggered}
     \end{subfigure}
     
     \begin{subfigure}[b]{\textwidth}
         \centering
         \caption{The mean number of \textit{detected} \texttt{Magma} bugs.}
         \includegraphics[width=\textwidth]{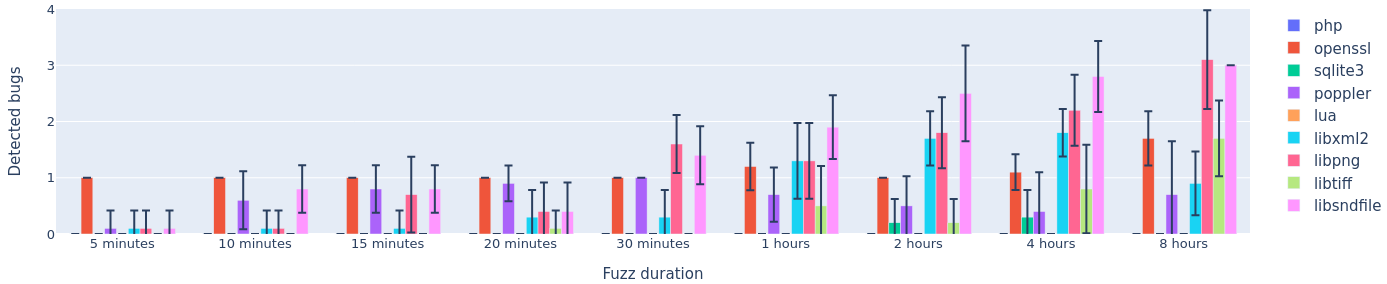}
         \label{detected}
     \end{subfigure}
     
    \caption{Fuzzing campaign duration versus the effectiveness of the fuzzing campaign.}
    \label{fig:duration}
\end{figure*}

Looking at the number of bugs \textit{reached}, Figure \ref{reached} shows no clear increase of bugs when the fuzzing campaign is allowed to last longer.
For some libraries like \texttt{sqlite3} and \texttt{libsndfile} the number of bugs does rise, while for others like \texttt{poppler} and \texttt{php} it declines.
While a decline in the number of bugs seems strange, this might be attributed to the stochastic nature of the fuzzing process.


Looking at the number of bugs \textit{triggered}, Figure \ref{triggered} shows that four out of nine libraries exhibit an increase in the number of bugs while increasing the fuzzing campaign duration.
For the library that exhibits the most significant change (\texttt{libsndfile}), the number of bugs has only doubled when the duration of the fuzzing campaign is thirty-two times its original duration.
For the other five out of nine libraries, there is no significant change to the number of bugs.


Looking at the number of bugs \textit{detected}, Figure \ref{detected} shows that five out of nine libraries show a positive change in the number of bugs when increasing the duration of the fuzzing campaigns.
Note, however, the difference in the scaling of the y-axes of Figures \ref{reached}, \ref{triggered}, and \ref{detected}.
Nonetheless, the difference of a fuzzer \textit{detecting} only one bug or three bugs is quite significant.

Both Google's \texttt{CIFuzz} and GitLab's coverage fuzzing implementations use 10 minute fuzzing durations as their default campaigns~\cite{GoogleIntegration,gitlab-cov-fuzz}.
This satisfies the desires of developers to have pipelines that take around 10 minutes~\cite{fow,Hilton2017Trade-offsFlexibility}, however in our view it would be beneficial to increase this default value.
From the results of the fuzzing campaign duration experiment, we find that selecting the 15 minute fuzz duration instead, this can lead to a more consistent effectiveness of the per-commit fuzzing campaigns, while still remaining quite close to the desires of developers.

\begin{mdframed}
In order to address \textbf{RQ.3}, we found that the gain in fuzzing effectiveness when increasing the time budget for \textit{per-commit} fuzzing campaigns is overall \textit{relatively poor}.
Campaigns of 15 minutes can be \textit{just as effective} as ones that take 8 hours, especially if lengthier campaigns are still \textit{regularly} used to fuzz snapshots of the repository.
\end{mdframed}

\section{Threats to Validity
\& Future Work
}\label{sec:6}
In this section, we discuss the limitations of our experimental study, along with several directions for future research.
Among these directions, we sketch a technique that could automatically determine the priority of commits, based on which a certain amount of resources can be proportionally allocated towards its fuzzing campaign.

\paragraph{Generality}
One of the points that could be made about the underlying software projects that are in use by the \texttt{Magma} benchmark, is that these libraries are written in C and quite far ahead in their development life cycle.
In turn, the experimental findings of this paper may not be representative for software projects that are written in other languages or for those that are still in their infancy stage, where most of the development still lies ahead.
However, we believe that the libraries included in our experiments represent a wide variety of popular software projects, and are therefore quite representative of the security-critical software repositories that would benefit most from continuous fuzzing practices.

\paragraph{Limited number of commits} 
In the target selection experiment, the number of processed commits varies greatly among the nine libraries and we processed a limited number of commits per library.
Ideally, all of the commits would be processed, however this is not achievable in practice without manually taking care of any problems that arise while processing the commits.
Also, only a chunk of the most recent commits are processed during the experiment, since those include the fuzzing harnesses, while earlier commits do not.
This may also affect the results of the experiment, since earlier commits probably contain more radical and a larger amount of changes than later commits.
In turn, using target selection in practice on software that is in its early development stages may achieve other results.

\paragraph{Less resources spent}
While conserving computational resources would be beneficial for the overall fuzzing effort, in some cases this may lead to bugs either being discovered later than would otherwise be the case, or - in the worst case - not being discovered at all.
When employing target selection in practice, there will be fuzzing campaigns that are skipped.
Although the corresponding commit may not have changed the fuzz target, it may be the case that the target did receive changes recently (by an older commit).
Fuzzing this target a few times (instead of only once) might reveal a bug that would otherwise not have been found yet.
Such cases can be covered by employing longer-running nightly fuzzing sessions that still fuzz all of the targets, catching the bugs that make it past the target selection approach.


\paragraph{Advanced target selection}
The PUT's behavior may be changed in ways other than the syntactic change of the program binary, e.g., via changes in the configuration.
In the future, we can employ more sophisticated forms of target selection.
Ideally, we would want to have a more robust approach that is more precise in determining the exact relation between commits and the available fuzz targets.

\paragraph{Directed continuous fuzzing}
When a more sophisticated form of target selection is implemented properly, we could further optimize the fuzzing process by actively steering the effort towards the actual code changes.
This is a technique known as directed greybox fuzzing, which was proposed by Böhme et al. \cite{bohme2017directed} in the form of a tool called \texttt{AFLGo}.
It would be quite interesting to investigate the impact of directed fuzzing in combination with advanced target selection on the effectiveness of continuous fuzzing practices, since this impact may be rather positive.
Directed fuzzing has not been part of the experiments presented in this work, since this form of fuzzing requires a specification of exactly which lines of code should be targeted within each fuzz target.
Seeing as the benchmark used in the experiments injects all of the bugs in a specific version of the target libraries, there would be a considerable bias when directing the fuzzers towards those code changes.
Thus, to investigate the effectiveness of directed fuzzing within continuous integration, alternative methods of evaluation would be necessary.
Moreover, to the best of our knowledge, continuous fuzzing in practice does not employ directed fuzzing either, which makes the setup in our experiments more realistic.

\paragraph{Automatic fuzzing priority}
Automatically determining the priority of a fuzzing campaign and allocating the appropriate amount of resources is another idea for future research.
\begin{mdframed}
The \textit{priority} of a fuzzing campaign could be determined automatically using several possible indicators like commit type, affected file types, commit message, and the size of the commit.
\end{mdframed}
Individual commits could be given a low priority, groups of commits a medium one, and merge requests a high one.
\texttt{HTML} files could be given a lower priority than \texttt{C} files.
Commit messages may give an indication of commit priority based on keywords that are included.
The size of a commit, whether measured in changed characters, changed lines of code, or in some other manner, may also indicate a certain priority.
Now that the fuzzing priority is determined, a certain amount of resources can be proportionally allocated towards the fuzzing campaign.
Low priorities can get the default (15 minutes) fuzz duration, medium ones the \textit{lunchtime} (1 hour) duration, and for high ones the \textit{bedtime} (8 hours) duration can be assigned.

\section{Related Work}\label{sec:7}


The tension between testing effectiveness and the compute cycles required in continuous testing practices has been of recent research interest~\cite{memon2017taming,RangnauBFT20,klooster2021effectiveness}. 
For instance, Memon et al.~\cite{memon2017taming} showed that testing each commit (coming every second on average) in CI is not sustainable and found out that only a small fraction of the test cases in Google's code base (only 63K among 5.5 Million test cases) had actually ever failed for a given period of commits. 

Böhme and Falk~\cite{Bohme2020} found that finding known bugs can be done in half the time with twice the computational resources.
However, they also showed that finding new bugs within the same time frame requires exponentially more resources.
This means that re-discovering vulnerabilities is cheap but finding new ones is expensive.

A recent study by Zhu and Böhme~\cite{Zhu2021RegressionFuzzing} on bug reports from OSS-Fuzz observed that the recent code changes (i.e. regressions) are responsible for 77 percent of bugs.
This means that the majority of bugs will be the result of code that is currently under development.
The authors then introduce regression greybox fuzzing in the form of a tool called \texttt{AFLChurn} as a way of focusing the fuzzing effort on code that was changed more recently or more often.
They propose to fuzz all commits at once, but to give code that is contained in commits more often or more recently, a higher priority for fuzzing.

Regression bugs, which are introduced by recent code changes, can still go undiscovered while using automatic reporting for 68 days on average, and 5 days on the median, as found by Zhu and Böhme \cite{Zhu2021RegressionFuzzing}.
This suggests that bugs of this type are quite hard to detect.
The authors also demonstrate that, only after a software project is well-fuzzed, additional fuzzing efforts need to be focused on recently changed code.

There is also recent work by Noller et al. \cite{NollerPBSNG20} that takes a step towards the detection of regression bugs in software that is under development.
The authors present a tool that combines feedback-directed greybox fuzzing with shadow symbolic execution to more efficiently explore and test the software under development.
The tool makes use of `divergence revealing heuristics based on resource consumption and control-flow information', in order to improve fuzzing efficiency.
Maximizing the divergence for fuzzer executions is done through the use of several differential metrics.

Most coverage-based fuzzing approaches keep any inputs that lead to parts of the code that were previously undiscovered, even though these parts may not have any impact on the security of the software under test.
The work by Wang et al. \cite{WangJLZBWS20} states that this is inefficient for vulnerability discovery since full coverage is hard to achieve within a reasonable amount of time, and the discovery of vulnerabilities needs to happen as early as possible.
The authors propose an approach that evaluates coverage by the impact on security, with metrics based on function, loop and basic block levels.
The fuzzer that they developed uses this approach to prioritize fuzzing inputs, and in doing so achieves better results in terms of more vulnerabilities found than other fuzzers.

\section{Conclusion}\label{sec:8}
In this work, we have investigated the effectiveness and scalability of continuous fuzzing techniques
.
Towards this end, we have conducted two distinct experiments for each fuzzing aspect respectively, by processing as many commits as possible from the live repositories of the considered open source libraries and by running over 1 CPU-year worth of fuzzing campaigns.

The experiments are based on a fuzzing benchmark that was recently proposed, featuring solid fuzzer metrics and a broad set of software libraries to evaluate fuzzers.
In order to make different fuzzers within an ensemble continuous fuzzing setup work together on the same commit, we found that a combination of fuzzing techniques results in a continuous fuzzing setup where different fuzzers can collaborate on the same targets effectively.
Therefore, we have extended the benchmark to support all of the aforementioned techniques, in order to properly evaluate continuous fuzzing campaigns.

The results of the first experiment suggests that even employing a simple technique for selecting the subset of fuzz targets which should be fuzzed, can result in saving more than half (55\% on average) of the computational resources needed for the continuous fuzzing effort.
We found that it suffices to only fuzz the targets of which the code was changed by the latest commit, in contrast to fuzzing all of the targets every time, when looking at per-commit fuzzing campaigns.

The second experiment shows that while having longer (e.g., 8 hours) fuzzing campaigns can boost the fuzzing effectiveness, employing per-commit fuzzing campaigns of 15 minutes in general strikes a solid balance between the desires of developers and the effectiveness of fuzzing.
We found that the gain in fuzzing effectiveness when increasing the time budget for per-commit fuzzing campaigns is overall relatively poor.
Campaigns of 15 minutes can be just as effective as ones that take 8 hours, especially if lengthier campaigns are still regularly used to fuzz snapshots of the repository.
Such lengthier fuzzing campaigns should always be executed when a pull request is ready to merge into the main branch, but when developing on a feature branch, each commit usually does not represent a mergeable state, and as such missing bugs with short fuzzing runs will not have an effect on the final product.

Finally, we sketch how the priority of a fuzzing campaign could be determined automatically by using several possible indicators like commit type, affected file types, commit message, and the size of the commit.
In turn, a certain amount of resources can be proportionally allocated towards each fuzzing campaign.



\bibliographystyle{acm}
\bibliography{main}

\end{document}